\documentclass[aps,10pt,twocolumn,
amsmath,showkeys,prl,superscriptaddress,notitlepage
]{revtex4-1}

\usepackage{dcolumn}
\usepackage{bm}
\usepackage[T1]{fontenc}
\usepackage{amsmath}
\usepackage{graphicx} 
\usepackage{tabularx}
\usepackage{multirow}
\usepackage{rotating} 
\usepackage{makecell}
\usepackage{amssymb}

\usepackage{hyperref}
\hypersetup{colorlinks=true, linkcolor=blue, citecolor=blue, urlcolor=blue}

\usepackage[normalem]{ulem}
\usepackage[usenames, dvipsnames]{xcolor}
\newcommand\blue[1]{{\color{blue}#1}}
\newcommand\red[1]{{\color{red}#1}}

\begin{document}
\title{
Magnetic competition in iron-based germanide and silicide superconductors}
\author{P. Villar Arribi}
\affiliation{European Synchrotron Radiation Facility, 71 Avenue des Martyrs, F-38000 Grenoble, France}
\affiliation{Laboratoire de Physique et Etude des Mat\'eriaux, UMR8213 CNRS/ESPCI/UPMC, Paris, France} 
\author{F. Bernardini}
\affiliation{Dipartimento di Fisica, Universit\`a di Cagliari, IT-09042 Monserrato, Italy}
\author{L. de' Medici}
\affiliation{Laboratoire de Physique et Etude des Mat\'eriaux, UMR8213 CNRS/ESPCI/UPMC, Paris, France} 
\author{P. Toulemonde}
\affiliation{Institut N\'eel, CNRS \& Univ. Grenoble Alpes, 38042 Grenoble, France}
\author{S. Tenc\'e}
\affiliation{CNRS, Univ. Bordeaux, ICMCB, UPR 9048, F-33600 Pessac, France}
\author{A. Cano}
\email{andres.cano@cnrs.fr}
\affiliation{Institut N\'eel, CNRS \& Univ. Grenoble Alpes, 38042 Grenoble, France}
\affiliation{Department of Materials, ETH Zurich, 8093 Zurich, Switzerland}
\date{\today}

\begin{abstract}
We address the ferromagnetic tendencies detrimental for superconductivity that are related to the substitution of the pnictogen As atom with Ge or Si, together with additional substitutions in the spacer layers in 122 and 1111 Fe-based superconductors.
Intermediate compounds in which these substitutions are realized individually are studied within density functional theory. We thus single out the control of spacer ions as an effective way to handle such a ferromagnetism, and we also show that it is suppressed in YFe$_2$Ge$_2$ under pressure ---which then can be expected to enhance its superconductivity.
\end{abstract}

\maketitle

Iron-based superconductors keep providing a very rich and intriguing platform for high-temperature unconventional superconductivity \cite{alloul-crp-16}. In these materials,
the Fe atom is invariably associated to pnictogen (As, P) or chalcogen (Se, Te, S) elements and, in practice, the most interesting superconducting properties are always obtained with either As or Se \cite{hosono-15}. 
The origin of this ``chemical'' limitation regarding alternative compounds remains unclear and attracts a research attention that is crucial for further advancing the field. 

Recently, this circumstance has been linked to the emergence of detrimental ferromagnetism as one goes from As/Se to the left in the periodic table \cite{valenti17} (see also \cite{valenti13}). However, there are two notable exceptions to this rule. Namely, the 122 germanide YFe$_2$Ge$_2$ with superconducting transition temperature $T_c \lesssim 1.8$~K \cite{chen-prl-16}, and the novel 1111 silicide hydride LaFeSiH displaying the second highest $T_c \simeq 11$~K among the 1111 parent compounds \cite{bernardini-prb-18} due to an unconventional mechanism \cite{yildirim18}.
In this paper, we examine how these intriguing Fe-based superconducting variants manage to run away from ferromagnetism. Specifically, we perform density functional theory (DFT) calculations and compare the resulting electronic structure and magnetic states with their closest pnictide counterparts. In doing so, we split the overall compositional change in two separate steps: changes in the ligands and changes in the spacer ions. This clarifies the competition between different magnetic instabilities, and enables the identification of fundamental design rules for the suppression of the ferromagnetic one that is necessary to promote superconductivity in novel Fe-based materials.

\begin{figure*}[tb!]
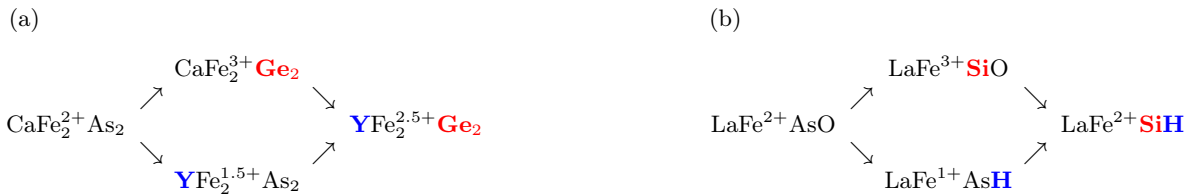

\flushleft \hspace{30pt} (a) \hspace{250pt}(b)
\begin{center}
CaFe$^{2+}_2$As$_2$ 
\begin{tabular}{ccc}
& CaFe$^{3+}_2$\red{{\bf Ge}$_2$} &\\
$\nearrow$ && $\searrow$
\\ \\
$\searrow$ && $\nearrow$ \\
& \blue{\bf Y}Fe$^{1.5+}_2$As$_2$ & \\
\end{tabular}
\blue{\bf Y}Fe$^{2.5+}_2$\red{{\bf Ge}$_2$} 
\hspace{80pt} LaFe$^{2+}$AsO 
\begin{tabular}{ccc}
& LaFe$^{3+}$\red{\bf Si}O &\\
$\nearrow$ && $\searrow$
\\ \\
$\searrow$ && $\nearrow$ \\
& LaFe$^{1+}$As\blue{\bf H} & \\
\end{tabular}
LaFe$^{2+}$\red{\bf Si}\blue{\bf H}
\end{center}
\caption{\label{fig:intermediate} (a) Scheme with the two possible compounds interpolating between the existing 122 superconductors CaFe$_2$As$_2$ and YFe$_2$Ge$_2$ that are considered in this work.
(b) Studied interpolation between the existing non-superconductor LaFeAsO and the superconductor LaFeSiH within the 1111 family. The oxidation states indicated in here represent nominal values in a simplified ionic-like picture.}
\end{figure*}

In the case of YFe$_2$Ge$_2$, the system can be seen as an hole-doped version of CaFe$_2$As$_2$ in its collapsed tetragonal phase \cite{chen-prl-16} where the tendency towards ferromagnetism is due to a Stoner instability \cite{singh-prb14,valenti17,subedi_prb14}. We exploit this connection and 
consider the intermediate compound CaFe$_2$Ge$_2$, reported for the first time in its pure parent phase in \cite{toulemonde}.
This novel compound interpolates the two previous 122 superconductors as illustrated in the upper path in Fig.\ref{fig:intermediate}(a), and is expected to be an even more hole-doped version of the reference CaFe$_2$As$_2$ compound, with a nominal oxidation state Fe$^{3+}$ of the iron atom under the assumption of ionic-like bonds (which has obvious limitations given the metallic character of the systems under consideration, in particular in the collapsed phases).
In addition, we also consider the alternative interpolation via the hypothetical compound YFe$_2$As$_2$ [lower path in Fig.\ref{fig:intermediate}(a)]. 
In this case, the intermediate compound represents an electron-doped version of the initial Ca pnictide since the nominal oxidation of the iron is reduced from Fe$^{2+}$ to Fe$^{1.5+}$ ({\it i.e.} an alternative to electron doping e.g. by replacement of Ca by La \cite{saha-prb11}).
These intermediate changes are under- and overcompensated respectively in the superconducting germanide YFe$_2$Ge$_2$, where the nominal oxidation of the iron becomes Fe$^{2.5+}$ again assuming a simplified ionic-like picture (as compared to Fe$^{2+}$ for the initial Ca pnictide).
In the case of the LaFeSiH superconductor we follow the same strategy and consider its interpolation to the reference LaFeAsO compound via the intermediate 1111 hypothetical materials LaFeSiO and LaFeAsH [Fig. \ref{fig:intermediate}(b)]. In this case, we have Fe$^{2+}$ in both LaFeAsO and LaFeSiH and hence the intermediate dopings are perfectly compensated instead (see Fig \ref{fig:intermediate}).     
The trends that emerge from the electronic and magnetic properties computed for these systems clearly show that, albeit the compounds with Ge/Si as ligands generally --but not always-- have a higher tendency to ferromagnetism compared to their pnictide counterparts, this tendency can be counteracted by the spacer ions which then allows superconductivity to emerge again.

\section{Methods}

Our DFT calculations are performed in the generalized gradient approximation (GGA) of Perdew, Burke and Ernzerhof~\cite{PBE} as implemented in {\sc{WIEN2k}}~\cite{WIEN2k}. 
Even if electronic correlations can play an important role in the paramagnetic phases of Fe-based superconductors, their strength is considerably reduced in the magnetic phases and tend to decrease with increasing magnetic polarization~\cite{Edelmann_BaCr2As2} which, in a first approximation, makes it possible the use of a DFT approach. This allows, in particular, a qualitative discussion on the magnetic tendencies and on the competition between different possible instabilities (see e.g. \cite{valenti15,valenti17}), as we do in this work. 
At the same time, GGA is known to overestimate the magnetism in the Fe-based superconductors (see e.g. \cite{mazin-prb08}). Thus, we performed complementary calculations within local density approximation (LDA) for the analysis according to the Stoner picture, which yield in fact a more clear trend. For a more quantitative discussion, however, an approach including the local many-body physics like dynamical mean-field theory should be used~\cite{Yin_kinetic_frustration_allFeSC}.

In our calculations, we use the lattice parameters and atomic positions reported in Ref.~\cite{Ca122_params} for CaFe$_2$As$_2$, which correspond to its tetragonal collapsed phase. For the novel germanide CaFe$_2$Ge$_2$ we use the lattice parameters and atomic positions measured experimentally (Table~\ref{139}) \cite{toulemonde}, while for YFe$_2$Ge$_2$ and the imaginary compound YFe$_2$As$_2$ we use the parameters reported in Ref.~\cite{Venturini}. We checked that our conclusions are robust with respect to reasonable variations of the relative Ge/As position. 

For the magnetic calculations we have selected muffin-tin radii of $R_{\rm MT}^{\mathrm{Y,Ca}}=2.50$ a.u.,  $R_{\rm MT}^{\mathrm{La}}=2.30$ a.u., $R_{\rm MT}^{\mathrm{Fe,Ge,Si,As}}=2.20$ a.u., and $R_{\rm MT}^{\mathrm{H}}=1.20$ a.u., and the same number of planewaves, which in {\sc{WIEN2k}} is set by the cutoff $R_{\rm MT}\cdot K_{max}=9.0$.
We have used 3 different magnetic supercells in order to accomodate all the possible magnetic structures and we have converged a $k$-mesh for each of them.
However, this introduces an error when comparing the energies of the different magnetic structures due to the finiteness of this $k$-mesh. We have estimated this error to be 6 meV.

We have considered the most relevant magnetic orders Fe-based superconductors, namely, the ferromagnetic (FM) order, an A-type order with FM Fe planes stacked antiferromagnetically along the $c$-axis, a single-stripe order with an in-plane arrangement that is FM along one direction and antiferromagnetic perpendicular to it, a double-stripe order with two lines of FM Fe moments that alternate antiferromagnetically in plane, and a checkerboard order with antiferromagnetic nearest Fe in plane.

For the study of YFe$_2$Ge$_2$ under pressure we have done structure optimizations for several values of pressure using VASP \cite{VASP} and the PAW pseudopotentials \cite{kresse99}. In these calculations we used a 400 eV plane-wave cutoff, a $15\times15\times15$ Monkhorst-Pack $k$-points mesh \cite{MP}, and a Gaussian smering of 0.1 eV. Y-4$s$ and Fe-3$p$ orbitals were explicitly included in the valence. The theoretical equilibrium structure was then used as input to {\sc{WIEN2k}} for spin-polarized calculations with ferromagnetic, single-, and double-stripe antiferromagnetic orders that we compare with the non-spin-polarized solution. In all these calculations we have used $R_{\rm MT}^{\mathrm{Y}}=2.50$ a.u., $R_{\rm MT}^{\mathrm{Fe}}=2.02$ a.u. and $R_{\rm MT}^{\mathrm{Ge}}=1.79$, $R_{\rm MT}\cdot K_{max}=9.0$ and the same converged $k$-mesh for each of the magnetic configurations.

\section{Results}
\subsection{YFe$_2$Ge$_2$ and related compounds}

\begin{figure*}[tb!]
\includegraphics[width=0.98\textwidth]{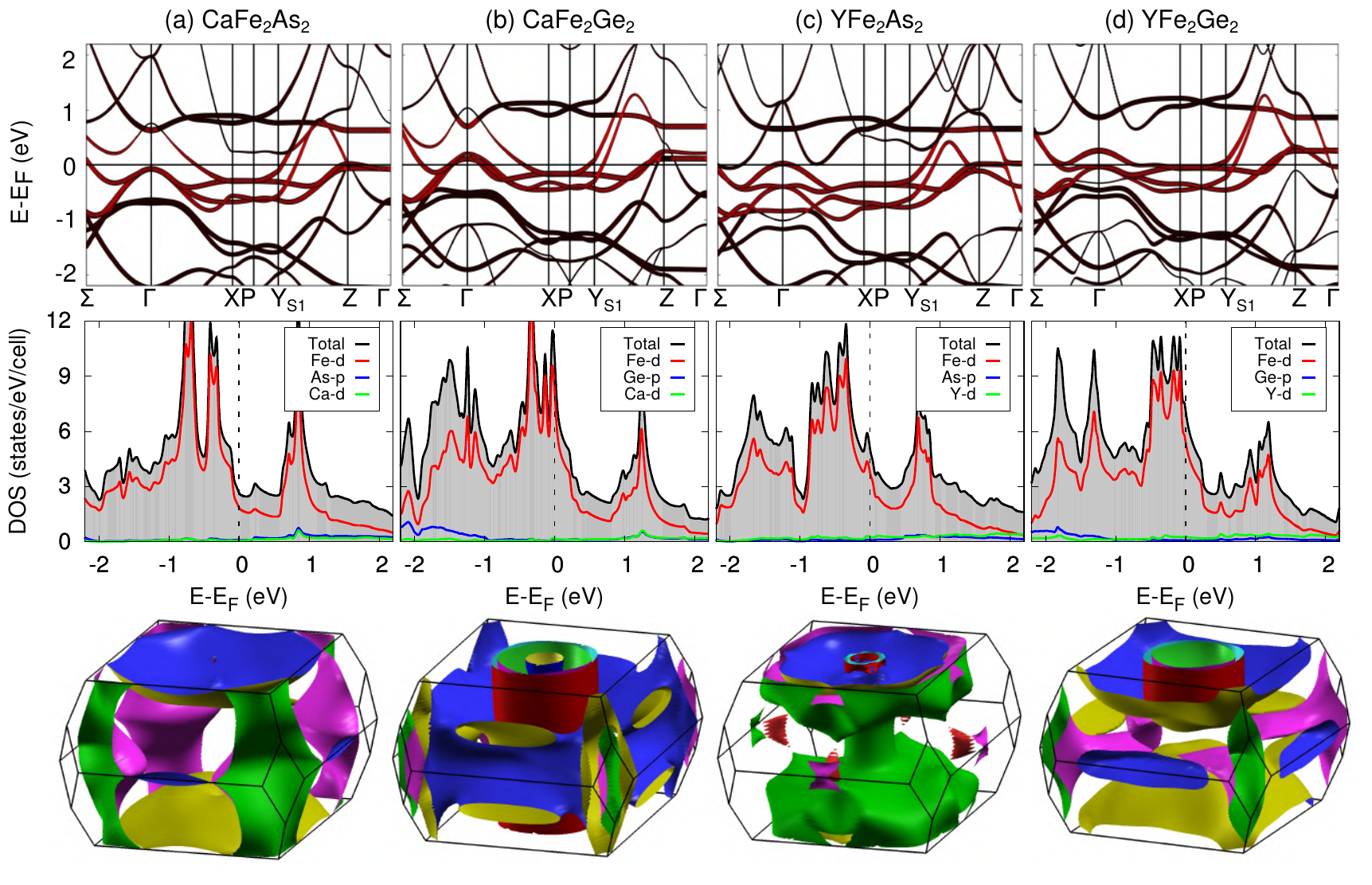}
\caption{\label{122_all} Results of the non-magnetic calculations for CaFe$_2$As$_2$ (a), CaFe$_2$Ge$_2$ (b), YFe$_2$As$_2$ (c) and YFe$_2$Ge$_2$ (d).
In each subfigure we show the electronic band structure with the Fe-d components of these bands and those crossing the Fermi level highlighted (upper panel), the density of states
(middle panel), and the Fermi surface (bottom panel) for each of the four compounds.}
\end{figure*}

We start by analyzing the relation between YFe$_2$Ge$_2$ and CaFe$_2$As$_2$ via the new compound CaFe$_2$Ge$_2$. This new 122 germanide crystalizes in the same tetragonal structure (space group $I4/mmm$) with the structural parameters summarized in Table \ref{139}. In particular, its lattice parameters perfectly match the direct extrapolation of the previous values obtained for the CaMn$_{2-x}$Fe$_x$Ge$_2$ series ($x \leq 1.9$) \cite{Welter03}, and they are very similar to those reported in YFe$_2$Ge$_2$~\cite{Venturini} and in the collapsed tetragonal phase of CaFe$_2$As$_2$~\cite{Ca122_params}. 

\begin{table}[b!]
\begin{tabular}{ccccc}
\multicolumn{5}{l}{
$I4/mmm$ (\#139)}\\ 
\multicolumn{5}{l}{$a=3.9922(6)$\AA{}, $c=10.702(2)$\AA{}}\\
\hline \hline 
 & Wyckoff pos.  & $x$ & $y$ & $z$ \\
\hline 
Ca & $2a$ & 0 & 0 & 0  \\
Fe & $4d$ & 0 & 1/2 & 1/4 \\
Ge & $4e$ & 0 & 0 & 0.3774(6) \\
\hline \hline
\end{tabular} 
\caption{Refined structural parameters of CaFe$_2$Ge$_2$ obtained from X-ray and electron diffraction at room temperature \cite{toulemonde}.}  
\label{139}
\end{table}

First, we computed the non-magnetic electronic structure of these compounds. 
The resulting band structure, density of states (DOS), and Fermi surface are summarized in Fig.~\ref{122_all}. 
Compared to CaFe$_2$As$_2$ and YFe$_2$Ge$_2$ (see also \cite{chen-prl-16,singh-prb14,valenti17,subedi_prb14}), the new compound CaFe$_2$Ge$_2$ displays very similar features at the Fermi energy with all 5 bands of mainly Fe-3$d$ character crossing the Fermi level. 
However, there is a shift upwards of these features that can be viewed as hole-doping in the iron plane.
Indeed this could have been anticipated from the fact that the nominal oxidation of the iron becomes Fe$^{3+}$ in the new system, compared to Fe$^{2+}$ in CaFe$_2$As$_2$ and Fe$^{2.5+}$ in YFe$_2$Ge$_2$. This extra doping in CaFe$_2$Ge$_2$ yields a substantial increase in the DOS at the Fermi level: from $\sim$ 2.5 eV$^{-1}$ in CaFe$_2$As$_2$ and $\sim$ 7.1 eV$^{-1}$ in YFe$_2$Ge$_2$ to $\sim$ 10.1 eV$^{-1}$ in CaFe$_2$Ge$_2$.

\begin{table}[t!] 
\footnotesize
\begin{tabular}{lcccc}
\hline \hline& {\bf CaFe$_2$As$_2$ }& {\bf CaFe$_2$Ge$_2$ } & {\bf YFe$_2$As$_2$ }& {\bf YFe$_2$Ge$_2$ } \\
\hline 
$I_\text{Fe}$ [eV] & 1.77 (2.01) & 1.53 (1.84) & 1.60 (2.80) & 1.68 (2.14) \\
$I$ [eV] & 0.405 (0.465) & 0.496 (0.611) & 0.438 (0.775) & 0.469 (0.611) \\
$N(0)$ [eV$^{-1}$] & 1.23 (1.23) & 4.89 (5.07) & 2.44 (2.46) & 3.50 (3.55) \\
$I  N(0)$  & 0.498 (0.572) & 2.42 (3.10) & 1.07 (1.91) & 1.64 (2.17) \\
\hline \hline & {\bf LaFeAsO }& {\bf LaFeSiO } & {\bf LaFeAsH }& {\bf LaFeSiH } \\
\hline 
$I_\text{Fe}$ [eV] & 1.26 (1.52) & 1.85 (2.35) & 2.28 (4.90) & 2.35 (2.97) \\
$I$ [eV]& 0.276 (0.341) & 0.402 (0.511) & 0.435 (0.834) & 0.426 (0.538) \\
$N(0)$ [eV$^{-1}$]& 2.24 (2.25) & 3.20 (3.20) & 5.67 (5.87) & 3.25 (3.24) \\
$I  N(0)$  & 0.618 (0.767) & 1.29 (1.64) & 2.47 (4.90) & 1.38 (1.74) \\
\hline \hline \\
\end{tabular}  
\caption{Parameters defining the Stoner criterion obtained from the fit of fixed-moment energy calculations to $E(m) = ({1 \over N(0)} - I){m^2\over 4} + b m^4$ at the LDA level (GGA values are in brackets). Here $N(0)$ is the paramagnetic density of states per unit cell and Fe spin at the Fermi level, $I$ is the Stoner parameter, and $m$ is the ferromagnetic Fe spin polarization. 
The individual Stoner parameter associated to the Fe is defined as $I_\text{Fe}= {I \over 2} ({N(0) \over N_{{\rm Fe}^{3d}}(0)  })^2$, where $N_{{\rm Fe}^{3d}}(0)$ is the partial DOS associated to the Fe 3$d$ orbitals (see \cite{Mazin-prb97}). 
\label{stoner}}

\end{table}

\begin{table}[b!] 
\begin{tabular}{cccccc}
\hline \hline
&\multicolumn{2}{c}{\bf\bf CaFe$_2$Ge$_2$} & \multicolumn{2}{c}{\bf YFe$_2$As$_2$} \\
 & \footnotesize{ $\Delta E$ (meV/Fe)} & \footnotesize{ $\mu_{\rm Fe}$($\mu_{\rm B}$)}  & \footnotesize{$\Delta E$ (meV/Fe)} &  \footnotesize{$\mu_{\rm Fe}$($\mu_{\rm B}$)}
 \\
\hline 
checkerboard & $-$99.2 & 1.78  & $-$ & $-$ \\
single-stripe & $-$112.5  & 1.61  &$-$ & $-$ \\
double-stripe & $-$127.4 & 1.74  &$-$ & $-$ \\
A-AFM & $-$116.5 & 1.33 & 133.8& 1.64 \\
FM & $-${\bf 132.6} & 1.33 & 106.5 & 1.64\\
\hline \hline 
\end{tabular} 
\caption{Energy difference per Fe atom (with respect to the non-spin-polarized calculation) and corresponding value of the Fe magnetic
moment for different magnetic orders in CaFe$_2$Ge$_2$ and YFe$_2$As$_2$. The energy of magnetic ground state is indicated in bold, and the orders for which the calculations did not converge by the symbol $-$. \label{mag-CaFe2Ge2}}
\end{table}

The increased DOS at the Fermi level in CaFe$_2$Ge$_2$ can lead to an enhanced ferromagnetic instability according to the Stoner picture (see e.g. Refs. \onlinecite{singh-prb14,valenti17}).
To confirm this, we performed fixed-moment calculations and computed the Stoner parameter $I$ from the linear response of the system \cite{Mazin-prb97}. The results are summarized in Table \ref{stoner}. We find that the total $I$, as well as the individual $I_{\rm Fe}$ of the Fe atom, shows a relatively weak dependence on the chemical substitution at the LDA level, staying virtually the same in our 122 systems. GGA, however, yields more apparent changes. In any case, the Stoner criterion $I N(0) > 1$ reveals a ferromagnetic trend for the set of 122 systems under consideration that can be safely related to the corresponding changes in the DOS.

To confirm this trend, we performed spin-polarized calculations for the ferromagnetic (FM) state and other relevant magnetic orders. The resulting energies and moments are summarized in Table~\ref{mag-CaFe2Ge2}. As we can see, the new germanide CaFe$_2$Ge$_2$ displays different local magnetic minima among which we do find a FM one.
In fact, the FM solution is obtained as the magnetic ground state in this system. This result is confirmed experimentally in Ref.~\onlinecite{toulemonde}.
Thus we see that the substitution of As with Ge does turn the striped antiferromagnetic CaFe$_2$As$_2$ into the ferromagnetic CaFe$_2$Ge$_2$, as anticipated from our previous DOS and Stoner analysis. The further substitution of Ca with Y reduces again the DOS at the Fermi level, which consequently reduces the FM tendency and favors its competition with other magnetic metastable orders~\cite{singh-prb14,subedi_prb14}. The end result is the non-magnetic YFe$_2$Ge$_2$, where superconductivity is possible again.

We can further check this insight by following the opposite order of chemical substitutions. That is, by changing first the spacer ions. Thus, we consider the intermediate imaginary compound YFe$_2$As$_2$. 
Fig.~\ref{122_all}(c) illustrates the electronic structure of this system obtained from non-spin-polarized calculations. 
The system is effectively an electron-doped version of CaFe$_2$As$_2$ in which, compared to that of CaFe$_2$Ge$_2$, the DOS at the Fermi level is suppressed down to 4.9 eV$^{-1}$. This suppression is expected to weaken the FM instability according to the Stoner criterion (see Table \ref{stoner}). In fact, the result of our spin-polarized calculations yields a non-magnetic ground state (see Table~\ref{mag-CaFe2Ge2}). Remarkably, among the considered solutions, only the FM and A-AFM (metastable) solutions are still realized in this case. 

This exercise confirms that the FM tendencies are inherently associated to the As $\to$ Ge (hole-doping) substitution in these 122 compounds according to a simple Stoner picture, and that instead the substitution Ca $\to$ Y (electron-doping) seems to have the opposite effect of suppressing them. We have validated this rationale considering in particular the new compound CaFe$_2$Ge$_2$ and the hypothetical system YFe$_2$As$_2$, both analyzed theoretically here for the first time.

\subsubsection{YFe$_2$Ge$_2$ under pressure}

\begin{figure}[b!]
\includegraphics[width=0.485\textwidth]{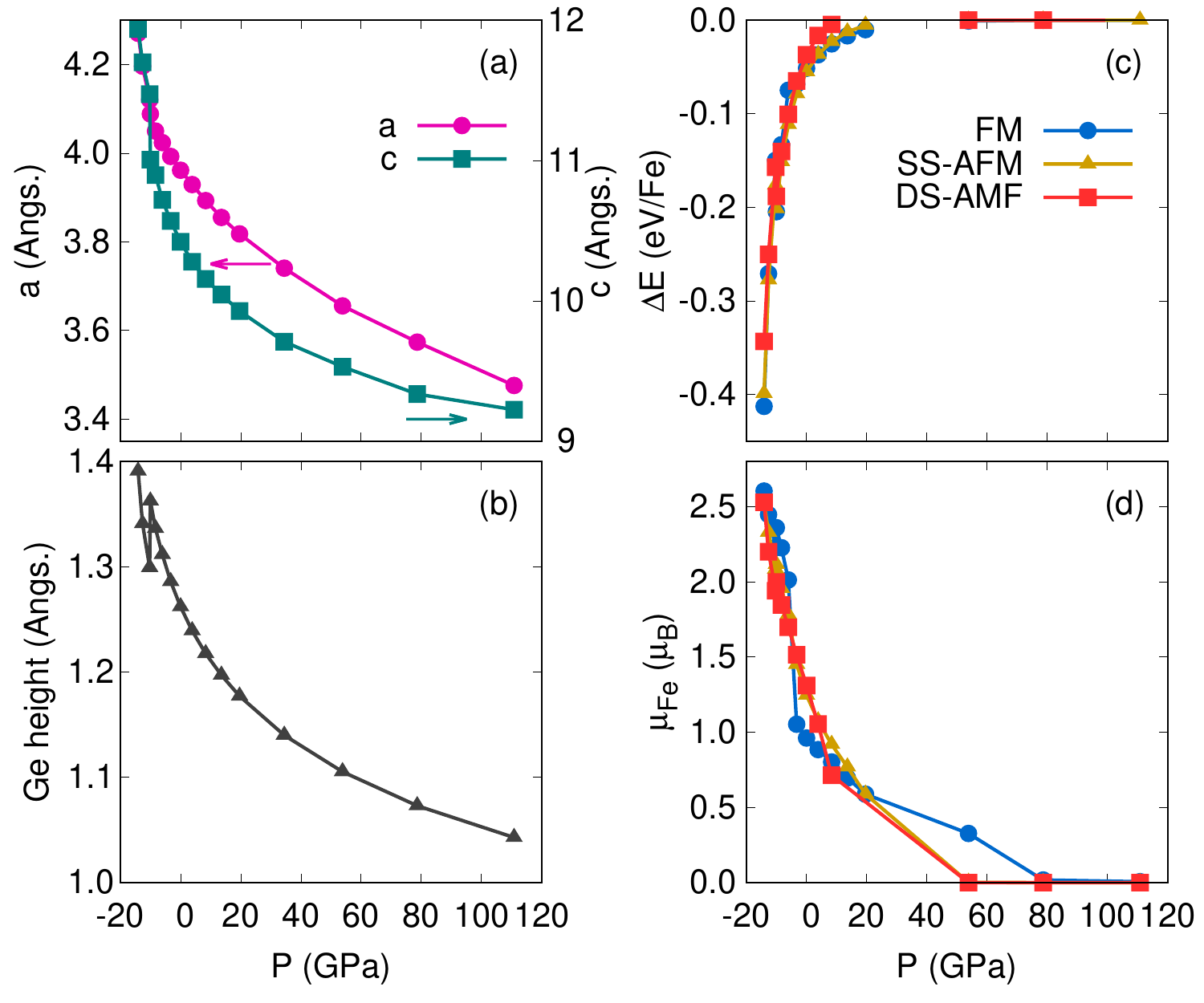}
\caption{\label{y122_pressure} 
Results for YFe$_2$Ge$_2$ as a function of hydrostatic pressure. (a) Lattice parameters $a$ (left y-axis) and $c$ (right y-axis). (b) Ge height with respect to the Fe-plane. (c) Energy difference per Fe atom of the different magnetic orders with respect to the non-magnetic solution. (d) Magnetic moment per Fe atom for the different magnetic configurations.
}
\end{figure}

The application of pressure is known to be effective in promoting superconductivity in the Fe-based superconductors, which is generally signaled by suppression of the competing magnetic phases. Thus, we have also performed a series of calculations for YFe$_2$Ge$_2$ under hydrostatic pressure. The main results are summarized in Fig.~\ref{y122_pressure}. 
Negative pressure produces an overall enhancement of the magnetic instabilities that promotes the ferromagnetic order as the ground state solution in very close proximity to the single- and double-stripe antiferromagnetic orders. 
The application of positive hydrostatic pressure, on the contrary, produces the suppression of these magnetic instabilities. 
This is clearly seen in Fig.~\ref{y122_pressure}, where energy difference between the paramagnetic and the magnetic states tends to zero by increasing the pressure and then is reversed up $\sim$ 60 GPa where the corresponding magnetic moments per Fe atom drop to zero. 
This confirms the vicinity of this system to a quantum critical point \cite{singh-prb14}, for which pressure is an effective control parameter enabling the general suppression of magnetism (not only the FM state). Thus one can speculate that, by tuning the distance to that special point using the external pressure, one can in principle enhance superconductivity in YFe$_2$Ge$_2$.

\subsection{LaFeSiH and related compounds}

\begin{figure*}[tb!]
\includegraphics[width=0.95\textwidth]{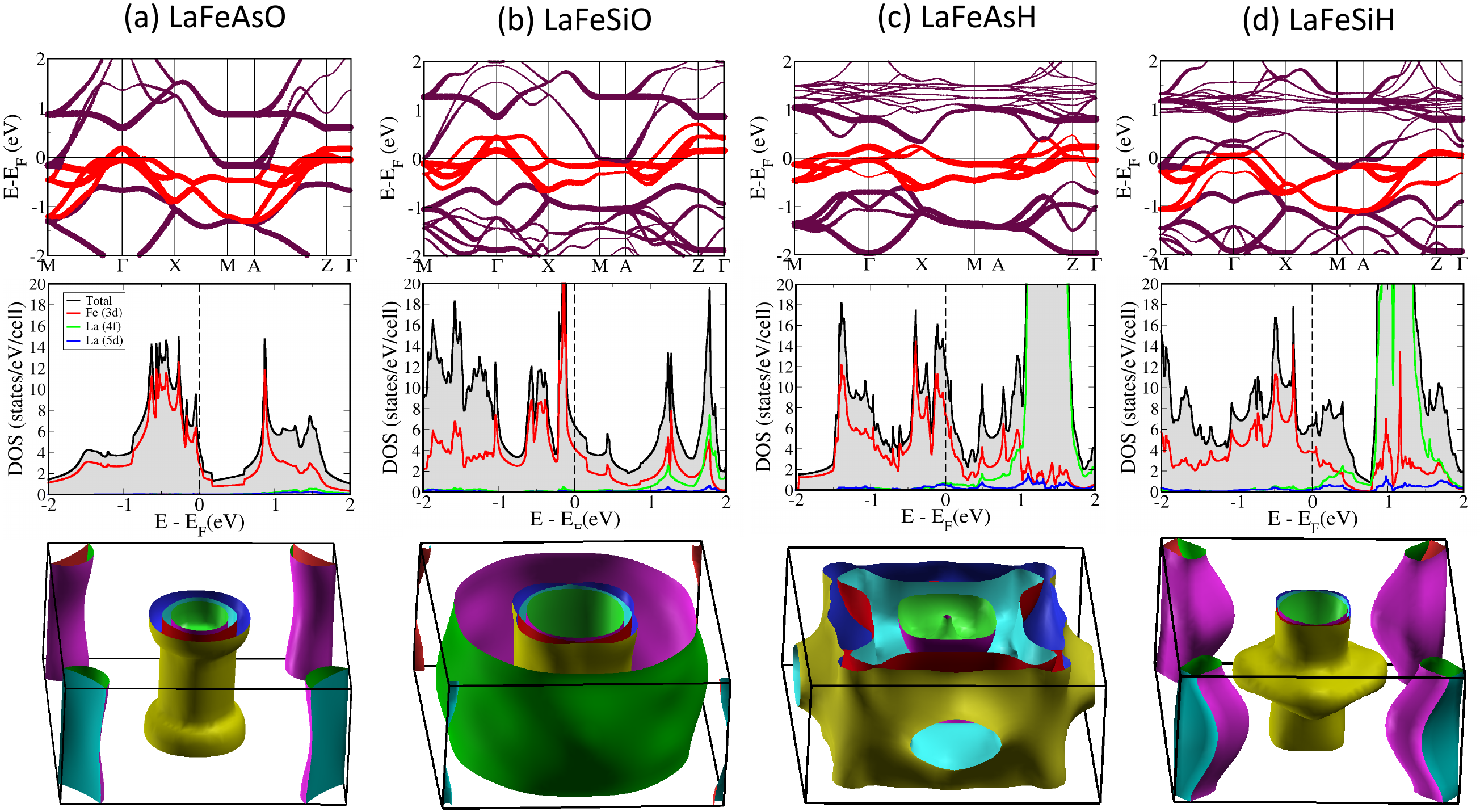}
\caption{\label{LaFeSiO-FS} Computed electronic band structure, DOS and Fermi surface for the different 1111 compounds discussed in the main text.}\end{figure*}

\begin{table*}[t!] 
\small
\begin{tabular}{ccccccccc}
\hline \hline&\multicolumn{2}{c}{\bf LaFeSiO } && \multicolumn{2}{c}{\bf LaFeAsH }&& \multicolumn{2}{c}{\bf LaFeSiH } \\
 & $\Delta E$ (meV/Fe) &  $\mu_{\rm Fe}$ ($\mu_{\rm B}$) && $\Delta E$ (meV/Fe) &  $\mu_{\rm Fe}$ ($\mu_{\rm B}$) && $\Delta E$ (meV/Fe) &  $\mu_{\rm Fe}$ ($\mu_{\rm B}$)\\
\hline 
checkerboard  & $-${\bf 47.50} & 1.38 && $-$ & $-$ && $-$5.71 & 0.90  \\
FM  & $-$40.25  & 1.11 && $-${\bf 265.8} & 2.61 && $-$11.11  & 0.65\\
double-stripe  & $-$41.65 & 1.33 && $-$218.5 & 2.20 && $-$11.26 & 1.04\\
single-stripe  & $-$16.00 & 0.96 && $-${\bf 261.6} & 2.15 && {\bf $-$44.56} & 1.16\\
\hline \hline \\
\end{tabular} 
\caption{Energy difference with respect to the non-spin-polarized calculation and corresponding value of the Fe magnetic moment
for different magnetic orders in LaFeSiO and LaFeAsH (this work) and LaFeSiH (from \cite{bernardini-prb-18}). The lowest energies for each compound are indicated in bold. \label{mag-LaFeSiH}}
\end{table*}

In the case of LaFeSiH we note that the system admits a FM solution even if the ground-state corresponds to the single-stripe one (see Table \ref{mag-LaFeSiH}) and the internal atomic positions have a non-negligible impact on it  (due to the weak character of such a FM tendency) \cite{yildirim18}.
This is in contrast to the reference arsenide LaFeAsO, in which the FM solution is absent \cite{yildirim08}. As in the previous section, in order to understand the FM tendency in the new 1111 silicide we consider the intermediate imaginary compound LaFeSiO in which only the As is replaced by the Si. That is, we follow the sequence of substitutions outlined in Fig.~\ref{fig:intermediate}(b).

From the charge point of view, the As $\to$ Si substitution increases the nominal oxidation of the iron from Fe$^{2+}$ to Fe$^{3+}$ and therefore
can be regarded as hole doping. In fact, as we can see in Fig. \ref{LaFeSiO-FS}, this substitution produces a rigid shift upwards of the band structure and DOS with respect to the Fermi level.
This shift, however, is quite substantial and drastically modifies the topology of the Fermi surface
(see Fig. \ref{LaFeSiO-FS}). Consequently, the complete As $\to$ Si substitution ends up into a hole over-doping that not only introduces a strong FM tendency, but also even changes the nature of the magnetic ground state from single-stripe to checkerboard antiferromagnetism as shown in Table \ref{mag-LaFeSiH}. However, we note that, even if the interpretation of the FM solution as due to a Stoner instability is still possible, now both the chemistry-induced changes in the DOS at the Fermi level and in the Stoner parameter play a role.
This is in contrast to the 122 systems analyzed previously and in \cite{valenti17}, where the FM tendency can be directly linked to the DOS alone. 
In any case, the initial electronic and magnetic features that are propitious for superconductivity are washed out in the case of the LaFeSiO intermediate compound.

Further on, the superconducting LaFeSiH compound implies the additional substitution of O for H. Again, from the charge point of view,
this substitution changes the nominal oxidation of the iron, which now goes from Fe$^{3+}$ back to Fe$^{2+}$. We then have an electron doping
that tends to compensate the hole doping introduced by the Si. In fact, the electronic band structure is shifted back and, importantly,
recovers the main features of the initial LaFeAsO (at the expense of displaying the additional La features closer to the Fermi level).
In particular, the topology of the Fermi surface is restored and the magnetic ground state is the single-stripe again \cite{bernardini-prb-18}.
The main conclusion of this analysis is that, analogously to the 122 case, the ions in the spacer layer can be used to restrain the FM tendencies induced by the substituted ligands (Si for As). Interestingly, the extra cation in the spacer layer of the 1111 structure represents an additional degree of freedom that can be used to this end. 

In order to clarify whether the enhanced FM is exclusively due to the As $\to$ Si substitution, we finally consider the hypothetical compound LaFeAsH [see Fig. \ref{fig:intermediate}(b)] with structural parameters directly extrapolated from \cite{hosono-srep16-LaFeAs(O-H)}. 
In this case, the nominal oxidation of the iron is reduced to Fe$^{1+}$ and therefore is expected to provide an extreme case of electron doping. According to the initial DOS of LaFeAsO shown in Fig. \ref{LaFeSiO-FS}(a) (middle panel) this should be safe in the sense that no FM should be promoted.    
However, as can be seen in Fig. \ref{LaFeSiO-FS}(c), the O $\to $ H substitution produces an important flattening of the bands rather than their rigid shift. In fact, compared to LaFeAsO, a similar flattening is also visible in LaFeSiO where it is superimposed to the shift upwards. 
In LaFeAsH, however, the flattening becomes dominant and so important that the interpretation in terms of simple charge doping breaks down. We note that the virtual crystal approximation employed in \cite{hosono-srep12} for the partial O $\to$ H substitution (up to 40\%) does not capture this effect, which has also been noticed for the 122 systems \cite{valenti17}. As a result, there is a substantial increase in overall DOS at the Fermi level to which La-5$d$ orbitals now also contribute together with a drastic change in the topology of the Fermi surface. Furthermore, compared with the non spin-polarized calculation, the magnetic solutions reduce more markedly the energy of the system with the FM and single-stripe ones effectively degenerate within the precision of our calculations (see Table \ref{mag-LaFeSiH}). This behavior, hardly expected from the reference LaFeAsO system, demonstrates that the modification of the spacer layer alone can also result into a strongly enhanced FM. Given the comparatively high value of the corresponding moment, the correct interpretation of the FM instability in LaFeAsH likely requires the extended Stoner theory \cite{Mazin-prb97,valenti17}, which is however out of the scope of this work. Beyond that, we note that the subsequent As $\to$ Si substitution can be seen as suppressing such a FM ---and thus enabling superconductivity in LaFeSiH--- which provides an interesting counterexample to our previous inferences and those in \cite{valenti17}. 
In the case of LaFeSiH, we also note that no FM solution is obtained in \cite{yildirim18} so that it is in fact rather sensitive to the internal atomic positions.

\section{Conclusions}

We have studied theoretically within the DFT framework two intermediate compounds that interpolate from CaFe$_2$As$_2$ to YFe$_2$Ge$_2$, two known Fe-based superconductors of the 122 family.
CaFe$_2$Ge$_2$, where only the As ligand (group V) is substituted by Ge (group IV) is found to be ferromagnetic. This can be understood within a Stoner picture as due to the strong enhancement of the DOS at the Fermi level in the paramagnetic phase. In contrast, YFe$_2$As$_2$ where only the cation in the spacer layer is substituted is predicted to be paramagnetic. The final superconductor YFe$_2$Ge$_2$ can thus be seen as collapsed version of CaFe$_2$As$_2$ where strong ferromagnetic tendencies induced by the substitution with Ge are mitigated by that with Y.
We have further confirmed the presence of a ``latent'' quantum critical point in the superconducting germanide and showed that it can be controlled by means of the external pressure. 
Thus, we speculate that the suppression of the residual FM tendencies associated to that point by the application of pressure can in principle enhance superconductivity in YFe$_2$Ge$_2$.

Analogously, in the 1111 family, we have studied the interpolation between LaFeAsO and the newly discovered superconductor LaFeSiH via the hypothetical compounds LaFeSiO and LaFeAsH. 
In contrast to LaFeAsO (where ferromagnetism is nonexistent) and LaFeSiH (where it is weak), both these hypothetical compounds display strong ferromagnetic tendencies. Thus, while LaFeSiO corroborates the trend formulated from the 122 systems, LaFeAsH provides an interesting counterexample in the sense that the subsequent substitution of As $\to$ Si to form the silicide LaFeSiH can be seen as weakening its ferromagnetism and hence enabling superconductivity. This, however, requires an important reconstruction of the electronic structure such that the main features of LaFeAsO emerge again.     

We have then concretely illustrated how ligands of the group IV generally ---but not always--- enhance the ferromagnetic tendencies by extending previous considerations \cite{valenti17} to newly discovered materials (i.e. the ferromagnetic CaFe$_2$Ge$_2$ and the superconducting LaFeSiH). We have shown, in particular, that the ions in the spacer layer ---and, to some extent, even the group-IV ligands themselves--- can be used to limit such a ferromagnetism in order to promote superconductivity in novel Fe-based compounds.

{\it Acknowledgements.---} P.V.A. and L.dM. are supported by the European Commission through the ERC-StG2016, StrongCoPhy4Energy, GA No724177. F.B. acknowledges partial support from the "Progetto biennale d'ateneo" UniCA/FdS/RAS CUP F72F16003050002. P.T., S.T, and A.C. are supported by the Grant ANR-18-CE30-0018-03 IRONMAN.

\bibliography{FeSivsFeGe}

\begin{thebibliography}{27}%
\makeatletter
\providecommand \@ifxundefined [1]{%
 \@ifx{#1\undefined}
}%
\providecommand \@ifnum [1]{%
 \ifnum #1\expandafter \@firstoftwo
 \else \expandafter \@secondoftwo
 \fi
}%
\providecommand \@ifx [1]{%
 \ifx #1\expandafter \@firstoftwo
 \else \expandafter \@secondoftwo
 \fi
}%
\providecommand \natexlab [1]{#1}%
\providecommand \enquote  [1]{``#1''}%
\providecommand \bibnamefont  [1]{#1}%
\providecommand \bibfnamefont [1]{#1}%
\providecommand \citenamefont [1]{#1}%
\providecommand \href@noop [0]{\@secondoftwo}%
\providecommand \href [0]{\begingroup \@sanitize@url \@href}%
\providecommand \@href[1]{\@@startlink{#1}\@@href}%
\providecommand \@@href[1]{\endgroup#1\@@endlink}%
\providecommand \@sanitize@url [0]{\catcode `\\12\catcode `\$12\catcode
  `\&12\catcode `\#12\catcode `\^12\catcode `\_12\catcode `\%12\relax}%
\providecommand \@@startlink[1]{}%
\providecommand \@@endlink[0]{}%
\providecommand \url  [0]{\begingroup\@sanitize@url \@url }%
\providecommand \@url [1]{\endgroup\@href {#1}{\urlprefix }}%
\providecommand \urlprefix  [0]{URL }%
\providecommand \Eprint [0]{\href }%
\providecommand \doibase [0]{http://dx.doi.org/}%
\providecommand \selectlanguage [0]{\@gobble}%
\providecommand \bibinfo  [0]{\@secondoftwo}%
\providecommand \bibfield  [0]{\@secondoftwo}%
\providecommand \translation [1]{[#1]}%
\providecommand \BibitemOpen [0]{}%
\providecommand \bibitemStop [0]{}%
\providecommand \bibitemNoStop [0]{.\EOS\space}%
\providecommand \EOS [0]{\spacefactor3000\relax}%
\providecommand \BibitemShut  [1]{\csname bibitem#1\endcsname}%
\let\auto@bib@innerbib\@empty
\bibitem [{\citenamefont {Alloul}\ and\ \citenamefont
  {Cano}(2016)}]{alloul-crp-16}%
  \BibitemOpen
  \bibfield  {author} {\bibinfo {author} {\bibfnamefont {H.}~\bibnamefont
  {Alloul}}\ and\ \bibinfo {author} {\bibfnamefont {A.}~\bibnamefont {Cano}},\
  }\href {\doibase https://doi.org/10.1016/j.crhy.2015.11.004} {\bibfield
  {journal} {\bibinfo  {journal} {Comptes Rendus Physique}\ }\textbf {\bibinfo
  {volume} {17}},\ \bibinfo {pages} {1 } (\bibinfo {year} {2016})},\ \bibinfo
  {note} {and the references therein.}\BibitemShut {Stop}%
\bibitem [{\citenamefont {Hosono}\ and\ \citenamefont
  {Kuroki}(2015)}]{hosono-15}%
  \BibitemOpen
  \bibfield  {author} {\bibinfo {author} {\bibfnamefont {H.}~\bibnamefont
  {Hosono}}\ and\ \bibinfo {author} {\bibfnamefont {K.}~\bibnamefont
  {Kuroki}},\ }\href {\doibase https://doi.org/10.1016/j.physc.2015.02.020}
  {\bibfield  {journal} {\bibinfo  {journal} {Physica C}\ }\textbf {\bibinfo
  {volume} {514}},\ \bibinfo {pages} {399 } (\bibinfo {year}
  {2015})}\BibitemShut {NoStop}%
\bibitem [{\citenamefont {Guterding}\ \emph {et~al.}(2017)\citenamefont
  {Guterding}, \citenamefont {Jeschke}, \citenamefont {Mazin}, \citenamefont
  {Glasbrenner}, \citenamefont {Bascones},\ and\ \citenamefont
  {Valent\'{\i}}}]{valenti17}%
  \BibitemOpen
  \bibfield  {author} {\bibinfo {author} {\bibfnamefont {D.}~\bibnamefont
  {Guterding}}, \bibinfo {author} {\bibfnamefont {H.~O.}\ \bibnamefont
  {Jeschke}}, \bibinfo {author} {\bibfnamefont {I.~I.}\ \bibnamefont {Mazin}},
  \bibinfo {author} {\bibfnamefont {J.~K.}\ \bibnamefont {Glasbrenner}},
  \bibinfo {author} {\bibfnamefont {E.}~\bibnamefont {Bascones}}, \ and\
  \bibinfo {author} {\bibfnamefont {R.}~\bibnamefont {Valent\'{\i}}},\ }\href
  {\doibase 10.1103/PhysRevLett.118.017204} {\bibfield  {journal} {\bibinfo
  {journal} {Phys. Rev. Lett.}\ }\textbf {\bibinfo {volume} {118}},\ \bibinfo
  {pages} {017204} (\bibinfo {year} {2017})}\BibitemShut {NoStop}%
\bibitem [{\citenamefont {Jeschke}\ \emph {et~al.}(2013)\citenamefont
  {Jeschke}, \citenamefont {Mazin},\ and\ \citenamefont
  {Valent\'{\i}}}]{valenti13}%
  \BibitemOpen
  \bibfield  {author} {\bibinfo {author} {\bibfnamefont {H.~O.}\ \bibnamefont
  {Jeschke}}, \bibinfo {author} {\bibfnamefont {I.~I.}\ \bibnamefont {Mazin}},
  \ and\ \bibinfo {author} {\bibfnamefont {R.}~\bibnamefont {Valent\'{\i}}},\
  }\href {\doibase 10.1103/PhysRevB.87.241105} {\bibfield  {journal} {\bibinfo
  {journal} {Phys. Rev. B}\ }\textbf {\bibinfo {volume} {87}},\ \bibinfo
  {pages} {241105} (\bibinfo {year} {2013})}\BibitemShut {NoStop}%
\bibitem [{\citenamefont {Chen}\ \emph {et~al.}(2016)\citenamefont {Chen},
  \citenamefont {Semeniuk}, \citenamefont {Feng}, \citenamefont {Reiss},
  \citenamefont {Brown}, \citenamefont {Zou}, \citenamefont {Logg},
  \citenamefont {Lampronti},\ and\ \citenamefont {Grosche}}]{chen-prl-16}%
  \BibitemOpen
  \bibfield  {author} {\bibinfo {author} {\bibfnamefont {J.}~\bibnamefont
  {Chen}}, \bibinfo {author} {\bibfnamefont {K.}~\bibnamefont {Semeniuk}},
  \bibinfo {author} {\bibfnamefont {Z.}~\bibnamefont {Feng}}, \bibinfo {author}
  {\bibfnamefont {P.}~\bibnamefont {Reiss}}, \bibinfo {author} {\bibfnamefont
  {P.}~\bibnamefont {Brown}}, \bibinfo {author} {\bibfnamefont
  {Y.}~\bibnamefont {Zou}}, \bibinfo {author} {\bibfnamefont {P.~W.}\
  \bibnamefont {Logg}}, \bibinfo {author} {\bibfnamefont {G.~I.}\ \bibnamefont
  {Lampronti}}, \ and\ \bibinfo {author} {\bibfnamefont {F.~M.}\ \bibnamefont
  {Grosche}},\ }\href {\doibase 10.1103/PhysRevLett.116.127001} {\bibfield
  {journal} {\bibinfo  {journal} {Phys. Rev. Lett.}\ }\textbf {\bibinfo
  {volume} {116}},\ \bibinfo {pages} {127001} (\bibinfo {year}
  {2016})}\BibitemShut {NoStop}%
\bibitem [{\citenamefont {Bernardini}\ \emph {et~al.}(2018)\citenamefont
  {Bernardini}, \citenamefont {Garbarino}, \citenamefont {Sulpice},
  \citenamefont {N\'u\~nez Regueiro}, \citenamefont {Gaudin}, \citenamefont
  {Chevalier}, \citenamefont {M\'easson}, \citenamefont {Cano},\ and\
  \citenamefont {Tenc\'e}}]{bernardini-prb-18}%
  \BibitemOpen
  \bibfield  {author} {\bibinfo {author} {\bibfnamefont {F.}~\bibnamefont
  {Bernardini}}, \bibinfo {author} {\bibfnamefont {G.}~\bibnamefont
  {Garbarino}}, \bibinfo {author} {\bibfnamefont {A.}~\bibnamefont {Sulpice}},
  \bibinfo {author} {\bibfnamefont {M.}~\bibnamefont {N\'u\~nez Regueiro}},
  \bibinfo {author} {\bibfnamefont {E.}~\bibnamefont {Gaudin}}, \bibinfo
  {author} {\bibfnamefont {B.}~\bibnamefont {Chevalier}}, \bibinfo {author}
  {\bibfnamefont {M.-A.}\ \bibnamefont {M\'easson}}, \bibinfo {author}
  {\bibfnamefont {A.}~\bibnamefont {Cano}}, \ and\ \bibinfo {author}
  {\bibfnamefont {S.}~\bibnamefont {Tenc\'e}},\ }\href {\doibase
  10.1103/PhysRevB.97.100504} {\bibfield  {journal} {\bibinfo  {journal} {Phys.
  Rev. B}\ }\textbf {\bibinfo {volume} {97}},\ \bibinfo {pages} {100504}
  (\bibinfo {year} {2018})}\BibitemShut {NoStop}%
\bibitem [{\citenamefont {Hung}\ and\ \citenamefont
  {Yildirim}(2018)}]{yildirim18}%
  \BibitemOpen
  \bibfield  {author} {\bibinfo {author} {\bibfnamefont {L.}~\bibnamefont
  {Hung}}\ and\ \bibinfo {author} {\bibfnamefont {T.}~\bibnamefont
  {Yildirim}},\ }\href {\doibase 10.1103/PhysRevB.97.224501} {\bibfield
  {journal} {\bibinfo  {journal} {Phys. Rev. B}\ }\textbf {\bibinfo {volume}
  {97}},\ \bibinfo {pages} {224501} (\bibinfo {year} {2018})}\BibitemShut
  {NoStop}%
\bibitem [{\citenamefont {Singh}(2014)}]{singh-prb14}%
  \BibitemOpen
  \bibfield  {author} {\bibinfo {author} {\bibfnamefont {D.~J.}\ \bibnamefont
  {Singh}},\ }\href {\doibase 10.1103/PhysRevB.89.024505} {\bibfield  {journal}
  {\bibinfo  {journal} {Phys. Rev. B}\ }\textbf {\bibinfo {volume} {89}},\
  \bibinfo {pages} {024505} (\bibinfo {year} {2014})}\BibitemShut {NoStop}%
\bibitem [{\citenamefont {Subedi}(2014)}]{subedi_prb14}%
  \BibitemOpen
  \bibfield  {author} {\bibinfo {author} {\bibfnamefont {A.}~\bibnamefont
  {Subedi}},\ }\href {\doibase 10.1103/PhysRevB.89.024504} {\bibfield
  {journal} {\bibinfo  {journal} {Phys. Rev. B}\ }\textbf {\bibinfo {volume}
  {89}},\ \bibinfo {pages} {024504} (\bibinfo {year} {2014})}\BibitemShut
  {NoStop}%
\bibitem [{\citenamefont {{P. Toulemonde, A. Sulpice, Ch. Lepoittevin, S.
  Pairis, S. Miraglia, and R. Haettel}}(tion)}]{toulemonde}%
  \BibitemOpen
  \bibfield  {author} {\bibinfo {author} {\bibnamefont {{P. Toulemonde, A.
  Sulpice, Ch. Lepoittevin, S. Pairis, S. Miraglia, and R. Haettel}}},\
  }\href@noop {} {\  (\bibinfo {year} {in preparation})}\BibitemShut {NoStop}%
\bibitem [{\citenamefont {Saha}\ \emph {et~al.}(2012)\citenamefont {Saha},
  \citenamefont {Butch}, \citenamefont {Drye}, \citenamefont {Magill},
  \citenamefont {Ziemak}, \citenamefont {Kirshenbaum}, \citenamefont {Zavalij},
  \citenamefont {Lynn},\ and\ \citenamefont {Paglione}}]{saha-prb11}%
  \BibitemOpen
  \bibfield  {author} {\bibinfo {author} {\bibfnamefont {S.~R.}\ \bibnamefont
  {Saha}}, \bibinfo {author} {\bibfnamefont {N.~P.}\ \bibnamefont {Butch}},
  \bibinfo {author} {\bibfnamefont {T.}~\bibnamefont {Drye}}, \bibinfo {author}
  {\bibfnamefont {J.}~\bibnamefont {Magill}}, \bibinfo {author} {\bibfnamefont
  {S.}~\bibnamefont {Ziemak}}, \bibinfo {author} {\bibfnamefont
  {K.}~\bibnamefont {Kirshenbaum}}, \bibinfo {author} {\bibfnamefont {P.~Y.}\
  \bibnamefont {Zavalij}}, \bibinfo {author} {\bibfnamefont {J.~W.}\
  \bibnamefont {Lynn}}, \ and\ \bibinfo {author} {\bibfnamefont
  {J.}~\bibnamefont {Paglione}},\ }\href {\doibase 10.1103/PhysRevB.85.024525}
  {\bibfield  {journal} {\bibinfo  {journal} {Phys. Rev. B}\ }\textbf {\bibinfo
  {volume} {85}},\ \bibinfo {pages} {024525} (\bibinfo {year}
  {2012})}\BibitemShut {NoStop}%
\bibitem [{\citenamefont {Perdew}\ \emph {et~al.}(1996)\citenamefont {Perdew},
  \citenamefont {Burke},\ and\ \citenamefont {Ernzerhof}}]{PBE}%
  \BibitemOpen
  \bibfield  {author} {\bibinfo {author} {\bibfnamefont {J.~P.}\ \bibnamefont
  {Perdew}}, \bibinfo {author} {\bibfnamefont {K.}~\bibnamefont {Burke}}, \
  and\ \bibinfo {author} {\bibfnamefont {M.}~\bibnamefont {Ernzerhof}},\ }\href
  {\doibase 10.1103/PhysRevLett.77.3865} {\bibfield  {journal} {\bibinfo
  {journal} {Physical Review Letters}\ }\textbf {\bibinfo {volume} {77}},\
  \bibinfo {pages} {3865} (\bibinfo {year} {1996})}\BibitemShut {NoStop}%
\bibitem [{\citenamefont {Schwarz}\ and\ \citenamefont {Blaha}(2003)}]{WIEN2k}%
  \BibitemOpen
  \bibfield  {author} {\bibinfo {author} {\bibfnamefont {K.}~\bibnamefont
  {Schwarz}}\ and\ \bibinfo {author} {\bibfnamefont {P.}~\bibnamefont
  {Blaha}},\ }\href {\doibase 10.1016/S0927-0256(03)00112-5} {\bibfield
  {journal} {\bibinfo  {journal} {Comp. Mater. Sci.}\ }\textbf {\bibinfo
  {volume} {28}},\ \bibinfo {pages} {259} (\bibinfo {year} {2003})}\BibitemShut
  {NoStop}%
\bibitem [{\citenamefont {Edelmann}\ \emph {et~al.}(2017)\citenamefont
  {Edelmann}, \citenamefont {Sangiovanni}, \citenamefont {Capone},\ and\
  \citenamefont {de' Medici}}]{Edelmann_BaCr2As2}%
  \BibitemOpen
  \bibfield  {author} {\bibinfo {author} {\bibfnamefont {M.}~\bibnamefont
  {Edelmann}}, \bibinfo {author} {\bibfnamefont {G.}~\bibnamefont
  {Sangiovanni}}, \bibinfo {author} {\bibfnamefont {M.}~\bibnamefont {Capone}},
  \ and\ \bibinfo {author} {\bibfnamefont {L.}~\bibnamefont {de' Medici}},\
  }\href {\doibase 10.1103/PhysRevB.95.205118} {\bibfield  {journal} {\bibinfo
  {journal} {Phys. Rev. B}\ }\textbf {\bibinfo {volume} {95}},\ \bibinfo
  {pages} {205118} (\bibinfo {year} {2017})}\BibitemShut {NoStop}%
\bibitem [{\citenamefont {{Glasbrenner}}\ \emph {et~al.}(2015)\citenamefont
  {{Glasbrenner}}, \citenamefont {{Mazin}}, \citenamefont {{Jeschke}},
  \citenamefont {{Hirschfeld}}, \citenamefont {{Fernandes}},\ and\
  \citenamefont {{Valent{\'{\i}}}}}]{valenti15}%
  \BibitemOpen
  \bibfield  {author} {\bibinfo {author} {\bibfnamefont {J.~K.}\ \bibnamefont
  {{Glasbrenner}}}, \bibinfo {author} {\bibfnamefont {I.~I.}\ \bibnamefont
  {{Mazin}}}, \bibinfo {author} {\bibfnamefont {H.~O.}\ \bibnamefont
  {{Jeschke}}}, \bibinfo {author} {\bibfnamefont {P.~J.}\ \bibnamefont
  {{Hirschfeld}}}, \bibinfo {author} {\bibfnamefont {R.~M.}\ \bibnamefont
  {{Fernandes}}}, \ and\ \bibinfo {author} {\bibfnamefont {R.}~\bibnamefont
  {{Valent{\'{\i}}}}},\ }\href {\doibase 10.1038/nphys3434} {\bibfield
  {journal} {\bibinfo  {journal} {Nature Physics}\ }\textbf {\bibinfo {volume}
  {11}},\ \bibinfo {pages} {953} (\bibinfo {year} {2015})}\BibitemShut
  {NoStop}%
\bibitem [{\citenamefont {Mazin}\ \emph {et~al.}(2008)\citenamefont {Mazin},
  \citenamefont {Johannes}, \citenamefont {Boeri}, \citenamefont {Koepernik},\
  and\ \citenamefont {Singh}}]{mazin-prb08}%
  \BibitemOpen
  \bibfield  {author} {\bibinfo {author} {\bibfnamefont {I.~I.}\ \bibnamefont
  {Mazin}}, \bibinfo {author} {\bibfnamefont {M.~D.}\ \bibnamefont {Johannes}},
  \bibinfo {author} {\bibfnamefont {L.}~\bibnamefont {Boeri}}, \bibinfo
  {author} {\bibfnamefont {K.}~\bibnamefont {Koepernik}}, \ and\ \bibinfo
  {author} {\bibfnamefont {D.~J.}\ \bibnamefont {Singh}},\ }\href {\doibase
  10.1103/PhysRevB.78.085104} {\bibfield  {journal} {\bibinfo  {journal} {Phys.
  Rev. B}\ }\textbf {\bibinfo {volume} {78}},\ \bibinfo {pages} {085104}
  (\bibinfo {year} {2008})}\BibitemShut {NoStop}%
\bibitem [{\citenamefont {Yin}\ \emph {et~al.}(2011)\citenamefont {Yin},
  \citenamefont {Haule},\ and\ \citenamefont
  {Kotliar}}]{Yin_kinetic_frustration_allFeSC}%
  \BibitemOpen
  \bibfield  {author} {\bibinfo {author} {\bibfnamefont {Z.~P.}\ \bibnamefont
  {Yin}}, \bibinfo {author} {\bibfnamefont {K.}~\bibnamefont {Haule}}, \ and\
  \bibinfo {author} {\bibfnamefont {G.}~\bibnamefont {Kotliar}},\ }\href
  {http://dx.doi.org/10.1038/nmat3120} {\bibfield  {journal} {\bibinfo
  {journal} {Nat Mater}\ }\textbf {\bibinfo {volume} {10}},\ \bibinfo {pages}
  {932} (\bibinfo {year} {2011})}\BibitemShut {NoStop}%
\bibitem [{\citenamefont {Kreyssig}\ \emph {et~al.}(2008)\citenamefont
  {Kreyssig}, \citenamefont {Green}, \citenamefont {Lee}, \citenamefont
  {Samolyuk}, \citenamefont {Zajdel}, \citenamefont {Lynn}, \citenamefont
  {Bud'ko}, \citenamefont {Torikachvili}, \citenamefont {Ni}, \citenamefont
  {Nandi}, \citenamefont {Le\~ao}, \citenamefont {Poulton}, \citenamefont
  {Argyriou}, \citenamefont {Harmon}, \citenamefont {McQueeney}, \citenamefont
  {Canfield},\ and\ \citenamefont {Goldman}}]{Ca122_params}%
  \BibitemOpen
  \bibfield  {author} {\bibinfo {author} {\bibfnamefont {A.}~\bibnamefont
  {Kreyssig}}, \bibinfo {author} {\bibfnamefont {M.~A.}\ \bibnamefont {Green}},
  \bibinfo {author} {\bibfnamefont {Y.}~\bibnamefont {Lee}}, \bibinfo {author}
  {\bibfnamefont {G.~D.}\ \bibnamefont {Samolyuk}}, \bibinfo {author}
  {\bibfnamefont {P.}~\bibnamefont {Zajdel}}, \bibinfo {author} {\bibfnamefont
  {J.~W.}\ \bibnamefont {Lynn}}, \bibinfo {author} {\bibfnamefont {S.~L.}\
  \bibnamefont {Bud'ko}}, \bibinfo {author} {\bibfnamefont {M.~S.}\
  \bibnamefont {Torikachvili}}, \bibinfo {author} {\bibfnamefont
  {N.}~\bibnamefont {Ni}}, \bibinfo {author} {\bibfnamefont {S.}~\bibnamefont
  {Nandi}}, \bibinfo {author} {\bibfnamefont {J.~B.}\ \bibnamefont {Le\~ao}},
  \bibinfo {author} {\bibfnamefont {S.~J.}\ \bibnamefont {Poulton}}, \bibinfo
  {author} {\bibfnamefont {D.~N.}\ \bibnamefont {Argyriou}}, \bibinfo {author}
  {\bibfnamefont {B.~N.}\ \bibnamefont {Harmon}}, \bibinfo {author}
  {\bibfnamefont {R.~J.}\ \bibnamefont {McQueeney}}, \bibinfo {author}
  {\bibfnamefont {P.~C.}\ \bibnamefont {Canfield}}, \ and\ \bibinfo {author}
  {\bibfnamefont {A.~I.}\ \bibnamefont {Goldman}},\ }\href {\doibase
  10.1103/PhysRevB.78.184517} {\bibfield  {journal} {\bibinfo  {journal} {Phys.
  Rev. B}\ }\textbf {\bibinfo {volume} {78}},\ \bibinfo {pages} {184517}
  (\bibinfo {year} {2008})}\BibitemShut {NoStop}%
\bibitem [{\citenamefont {Venturini}\ and\ \citenamefont
  {Malaman}(1996)}]{Venturini}%
  \BibitemOpen
  \bibfield  {author} {\bibinfo {author} {\bibfnamefont {G.}~\bibnamefont
  {Venturini}}\ and\ \bibinfo {author} {\bibfnamefont {B.}~\bibnamefont
  {Malaman}},\ }\href {\doibase https://doi.org/10.1016/0925-8388(95)02140-X}
  {\bibfield  {journal} {\bibinfo  {journal} {Journal of Alloys and Compounds}\
  }\textbf {\bibinfo {volume} {235}},\ \bibinfo {pages} {201 } (\bibinfo {year}
  {1996})}\BibitemShut {NoStop}%
\bibitem [{\citenamefont {Kresse}\ and\ \citenamefont {Hafner}(1993)}]{VASP}%
  \BibitemOpen
  \bibfield  {author} {\bibinfo {author} {\bibfnamefont {G.}~\bibnamefont
  {Kresse}}\ and\ \bibinfo {author} {\bibfnamefont {J.}~\bibnamefont
  {Hafner}},\ }\href {\doibase 10.1103/PhysRevB.47.558} {\bibfield  {journal}
  {\bibinfo  {journal} {Phys. Rev. B}\ }\textbf {\bibinfo {volume} {47}},\
  \bibinfo {pages} {558} (\bibinfo {year} {1993})}\BibitemShut {NoStop}%
\bibitem [{\citenamefont {Kresse}\ and\ \citenamefont
  {Joubert}(1999)}]{kresse99}%
  \BibitemOpen
  \bibfield  {author} {\bibinfo {author} {\bibfnamefont {G.}~\bibnamefont
  {Kresse}}\ and\ \bibinfo {author} {\bibfnamefont {D.}~\bibnamefont
  {Joubert}},\ }\href {\doibase 10.1103/PhysRevB.59.1758} {\bibfield  {journal}
  {\bibinfo  {journal} {Phys. Rev. B}\ }\textbf {\bibinfo {volume} {59}},\
  \bibinfo {pages} {1758} (\bibinfo {year} {1999})}\BibitemShut {NoStop}%
\bibitem [{\citenamefont {Monkhorst}\ and\ \citenamefont {Pack}(1976)}]{MP}%
  \BibitemOpen
  \bibfield  {author} {\bibinfo {author} {\bibfnamefont {H.~J.}\ \bibnamefont
  {Monkhorst}}\ and\ \bibinfo {author} {\bibfnamefont {J.~D.}\ \bibnamefont
  {Pack}},\ }\href {\doibase 10.1103/PhysRevB.13.5188} {\bibfield  {journal}
  {\bibinfo  {journal} {Phys. Rev. B}\ }\textbf {\bibinfo {volume} {13}},\
  \bibinfo {pages} {5188} (\bibinfo {year} {1976})}\BibitemShut {NoStop}%
\bibitem [{\citenamefont {Welter}\ and\ \citenamefont
  {Malaman}(2003)}]{Welter03}%
  \BibitemOpen
  \bibfield  {author} {\bibinfo {author} {\bibfnamefont {R.}~\bibnamefont
  {Welter}}\ and\ \bibinfo {author} {\bibfnamefont {B.}~\bibnamefont
  {Malaman}},\ }\href {\doibase https://doi.org/10.1016/S0925-8388(02)01354-3}
  {\bibfield  {journal} {\bibinfo  {journal} {Journal of Alloys and Compounds}\
  }\textbf {\bibinfo {volume} {354}},\ \bibinfo {pages} {35 } (\bibinfo {year}
  {2003})}\BibitemShut {NoStop}%
\bibitem [{\citenamefont {Mazin}\ and\ \citenamefont
  {Singh}(1997)}]{Mazin-prb97}%
  \BibitemOpen
  \bibfield  {author} {\bibinfo {author} {\bibfnamefont {I.~I.}\ \bibnamefont
  {Mazin}}\ and\ \bibinfo {author} {\bibfnamefont {D.~J.}\ \bibnamefont
  {Singh}},\ }\href {\doibase 10.1103/PhysRevB.56.2556} {\bibfield  {journal}
  {\bibinfo  {journal} {Phys. Rev. B}\ }\textbf {\bibinfo {volume} {56}},\
  \bibinfo {pages} {2556} (\bibinfo {year} {1997})}\BibitemShut {NoStop}%
\bibitem [{\citenamefont {Yildirim}(2008)}]{yildirim08}%
  \BibitemOpen
  \bibfield  {author} {\bibinfo {author} {\bibfnamefont {T.}~\bibnamefont
  {Yildirim}},\ }\href {\doibase 10.1103/PhysRevLett.101.057010} {\bibfield
  {journal} {\bibinfo  {journal} {Phys. Rev. Lett.}\ }\textbf {\bibinfo
  {volume} {101}},\ \bibinfo {pages} {057010} (\bibinfo {year}
  {2008})}\BibitemShut {NoStop}%
\bibitem [{\citenamefont {Kobayashi}\ \emph {et~al.}(2016)\citenamefont
  {Kobayashi}, \citenamefont {Yamaura}, \citenamefont {Iimura}, \citenamefont
  {Maki}, \citenamefont {Sagayama}, \citenamefont {Kumai}, \citenamefont
  {Murakami}, \citenamefont {Takahashi}, \citenamefont {Matsuishi},\ and\
  \citenamefont {Hosono}}]{hosono-srep16-LaFeAs(O-H)}%
  \BibitemOpen
  \bibfield  {author} {\bibinfo {author} {\bibfnamefont {K.}~\bibnamefont
  {Kobayashi}}, \bibinfo {author} {\bibfnamefont {J.-i.}\ \bibnamefont
  {Yamaura}}, \bibinfo {author} {\bibfnamefont {S.}~\bibnamefont {Iimura}},
  \bibinfo {author} {\bibfnamefont {S.}~\bibnamefont {Maki}}, \bibinfo {author}
  {\bibfnamefont {H.}~\bibnamefont {Sagayama}}, \bibinfo {author}
  {\bibfnamefont {R.}~\bibnamefont {Kumai}}, \bibinfo {author} {\bibfnamefont
  {Y.}~\bibnamefont {Murakami}}, \bibinfo {author} {\bibfnamefont
  {H.}~\bibnamefont {Takahashi}}, \bibinfo {author} {\bibfnamefont
  {S.}~\bibnamefont {Matsuishi}}, \ and\ \bibinfo {author} {\bibfnamefont
  {H.}~\bibnamefont {Hosono}},\ }\href {https://doi.org/10.1038/srep39646}
  {\bibfield  {journal} {\bibinfo  {journal} {Scientific Reports}\ }\textbf
  {\bibinfo {volume} {6}},\ \bibinfo {pages} {39646 EP } (\bibinfo {year}
  {2016})}\BibitemShut {NoStop}%
\bibitem [{\citenamefont {Iimura}\ \emph {et~al.}(2012)\citenamefont {Iimura},
  \citenamefont {Matsuishi}, \citenamefont {Sato}, \citenamefont {Hanna},
  \citenamefont {Muraba}, \citenamefont {Kim}, \citenamefont {Kim},
  \citenamefont {Takata},\ and\ \citenamefont {Hosono}}]{hosono-srep12}%
  \BibitemOpen
  \bibfield  {author} {\bibinfo {author} {\bibfnamefont {S.}~\bibnamefont
  {Iimura}}, \bibinfo {author} {\bibfnamefont {S.}~\bibnamefont {Matsuishi}},
  \bibinfo {author} {\bibfnamefont {H.}~\bibnamefont {Sato}}, \bibinfo {author}
  {\bibfnamefont {T.}~\bibnamefont {Hanna}}, \bibinfo {author} {\bibfnamefont
  {Y.}~\bibnamefont {Muraba}}, \bibinfo {author} {\bibfnamefont {S.~W.}\
  \bibnamefont {Kim}}, \bibinfo {author} {\bibfnamefont {J.~E.}\ \bibnamefont
  {Kim}}, \bibinfo {author} {\bibfnamefont {M.}~\bibnamefont {Takata}}, \ and\
  \bibinfo {author} {\bibfnamefont {H.}~\bibnamefont {Hosono}},\ }\href
  {https://doi.org/10.1038/ncomms1913} {\bibfield  {journal} {\bibinfo
  {journal} {Nature Communications}\ }\textbf {\bibinfo {volume} {3}},\
  \bibinfo {pages} {943 EP } (\bibinfo {year} {2012})}\BibitemShut {NoStop}%
\end{thebibliography}%

\end{document}